\documentclass[english,aps,twocolumn]{revtex4}
\usepackage[T1]{fontenc}
\usepackage[latin9]{inputenc}
\usepackage{float}
\usepackage{graphicx}
\usepackage{amssymb}
\usepackage{babel}

\begin{document}

\title{Vibrational wave packet induced oscillations in two-dimensional electronic spectra.\\ I. Experiments}

\author{Alexandra Nemeth$^{1,*}$, Franz Milota$^2$, Tom\'{a}\v{s} Man\v{c}al$^3$, Vladim\'ir Luke\v{s}$^4$,
J\"{u}rgen Hauer$^1$, Harald F. Kauffmann$^{5}$, Jaroslaw Sperling$^{6}$}

\affiliation{$^1$ Faculty of Chemistry, University of
Vienna, W\"{a}hringerstrasse 42, 1090 Vienna, Austria\\
$^2$ Lehrstuhl f�r BioMolekulare Optik, Ludwig-Maximilians-Universit�t, Oettingenstra�e 67, 80538 M�nchen, Germany\\
$^3$ Institute of Physics, Faculty of Mathematics and Physics,
Charles University, Ke Karlovu 5, Prague, 121 16 Czech Republic\\
$^4$ Department of Chemical Physics, Slovak Technical University,
Radlinsk\'eho 9, 81237 Bratislava, Slovakia\\
$^5$ Ultrafast Dynamics Group, Faculty of Physics, Vienna University of Technology, Wiedner
Hauptstrasse 8 - 10, 1040 Vienna, Austria\\
$^6$ Newport Spectra-Physics, Guerickeweg 7, 64291 Darmstadt, Germany}

\vspace{2cm}

\begin{abstract}
This is the first in a series of two papers investigating the effect
of electron-phonon coupling in two-dimensional Fourier transformed
electronic spectroscopy. We present a series of one- and
two-dimensional nonlinear spectroscopic techniques for studying a
dye molecule in solution. Ultrafast laser pulse excitation of an
electronic transition coupled to vibrational modes induces a
propagating vibrational wave packet that manifests itself in
oscillating signal intensities and line-shapes. For the
two-dimensional electronic spectra we can attribute the observed
modulations to periodic enhancement and decrement of the relative
amplitudes of rephasing and non-rephasing contributions to the total
response. Different metrics of the two-dimensional signals are shown
to relate to the frequency-frequency correlation function which
provides the connection between experimentally accessible
observations and the underlying microscopic molecular dynamics. A
detailed theory of the time-dependent two-dimensional spectral
line-shapes is presented in the accompanying paper [T. Man\v{c}al
\emph{et al.}, arXiv:1003.xxxx].

\vspace{0.5cm}\noindent $^{*}$ alexandra.nemeth@univie.ac.at
\end{abstract}

\maketitle

\section{Introduction}

Coherent wave packet generation in the manifolds of molecular exited
states takes place if the excitation pulses are spectrally broad
enough to coherently excite a number of vibrational or electronic
levels. Femtosecond (fs) laser pulses meet this criterion in a
number of molecular systems and experiments employing such pulses
often show signatures of wave packet motion. Modulations of signal
intensities induced by vibrational wave packet motion have been
observed in a number of nonlinear spectroscopic experiments,
including pump-probe (PP) \cite{Zewail1993, FragnitoShank1989,
KobayashiOhtani2001}, transient grating (TG)
\cite{LarsenFleming2001}, and a variety of photon-echo
spectroscopies \cite{BeckerShank1989, SchoenleinShank1993,
LarsenFleming2001, deBoeijWiersmaI1998, deBoeijWiersmaI1996}. A
number of theoretical investigations supported the conclusions drawn
in these experiments \cite{YanMukamel1990, PollardMathies1990,
PollardMathies1992, DharNelson1994, OhtaFlemingI2001}. The influence
of intramolecular vibrations should be expected consequently also in
the recently implemented heterodyne-detected two-dimensional
electronic spectroscopy (2D-ES) technique \cite{HyblJonas1998,
TianWarren2003, CowanMiller2004, BrixnerFlemingI2004,
BorcaCundiff2005, GundogduNelson2007, NemethSperlingI2008,
MilotaKauffmannII2009}, which is based on the correlation of
electronic coherences evolving during two time periods. 2D
electronic spectra recorded for distinct waiting times provide
information on coupling patterns and evolution of populations and
coherences via line-shape modulations and the appearance and
disappearance of cross peaks \cite{HyblJonas2002, TianWarren2003,
BorcaCundiff2005, BrixnerFleming2005, StiopkinFleming2006,
NemethSperlingI2008, NemethMancal2009}.

Recent theoretical works on model dimer \cite{PisliakovFleming2006,
ChengFleming2008} and trimer \cite{ChengFleming2007} systems, and
experimental investigations on excitonic complexes \cite{EngelFlemingI2007, MilotaKauffmann2009}
and conjugated polymers \cite{ColliniScholes2009} reported on
signatures of electronic wave packets in 2D electronic spectra
resulting from coherent excitation of several electronic levels.
Possible enhancement of the energy transfer efficiency in
photosynthetic light harvesting complexes, due to the presence of
excitonic wave packets, was suggested by Engel \emph{et al.}
\cite{EngelFlemingI2007}. An assignment of the observed spectral
modulations to an excitonic wave packet motion was made, based on a
good agreement of the experiment with the predictions of an
excitonic model \cite{PisliakovFleming2006, EngelFlemingI2007}. In the general case,
however, contributions of vibrational wave packet motion cannot be
excluded and it is therefore of high importance to study the
vibrational case separately. To the best of our knowledge,
vibrational effects in 2D-ES have been addressed so far only
theoretically \cite{GallagherJonas1999, EgorovaDomckeI2007} and in our
recent work on the same molecular system \cite{NemethSperlingI2008}.

\begin{figure}
\begin{center}
\includegraphics{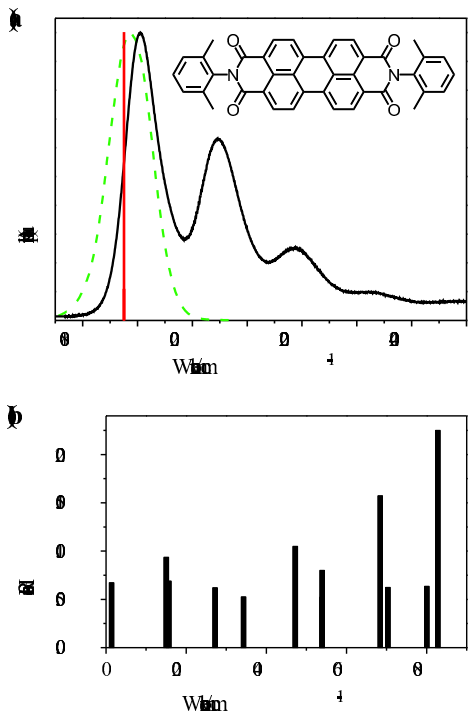}
\caption{(a) Linear absorption spectrum of PERY in toluene (black
solid line) and spectrum of the excitation pulses (green dashed
line). The red line indicates the position of the lowest energy
transition obtained from ZINDO/S calculations. The inset shows the
chemical structure of PERY. (b) Huang-Rhys factors determined from
semiemprical AM1 calculations. Only modes that posses Huang-Rhys
factors larger than 0.005 are plotted.}\label{Fig_LA}
\end{center}
\end{figure}

To study the effect of vibrational coherence without perturbations
from additional electronic levels, a two-level system is
preferential. Quantum-chemistry calculations (cf. Section II)
indicate that
N,N'-bis(2,6-dimethylphenyl)perylene-3,4,9,10-tetracarboxylicdiimide
(PERY), when excited at 20000~cm$^{-1}$, can be approximated as a
two-level electronic system with a number of vibrational modes
coupled to the electronic transition \cite{LarsenFleming1999}. In
the linear absorption spectrum of PERY in solution four well
resolved vibrational peaks with a frequency separation of
$\approx$1400~cm$^{-1}$ can be discerned (Fig.~\ref{Fig_LA}a). In
contrast to a recent work by Tekavec \emph{et al.}
\cite{TekavecOgilvie2009}, in which white light pulses where used to
probe the vibronic dynamics, this progression can not be covered
with the bandwidth of our laser pulses (900~cm$^{-1}$). However, by
tuning the central excitation energy approximately to the center of
the lowest energy peak, nonlinear spectroscopic signals such as
transient grating (TG) and three-pulse photon echo peak shift
(3PEPS) have been shown to be modulated by a low frequency
(140~cm$^{-1}$) vibrational mode \cite{LarsenFleming1999}.

In this contribution we investigate experimentally how coherent
excitation of several vibrational levels influences 2D electronic
correlation and relaxation spectra. To this end 2D electronic
spectra of PERY have been recorded for 20 waiting times $t_2$ ranging from 0
to 800~fs covering several periods of pronounced vibrational
modulations of the spectra. In a recent work on the same molecular
system we have shown that this low frequency mode induces a periodic
change in the ellipticity of the real (absorptive) part and in the
slope of the nodal line separating the positive and negative
contribution of the imaginary (dispersive) part
\cite{NemethSperlingI2008}. Building up on these qualitative
observations, in this contribution we will assign these modulations
to periodic enhancement and decrement of the rephasing and
non-rephasing signal parts and establish connections to the
frequency-frequency correlation function M(t). In the accompanying
paper Ref.~\cite{MancalSperling2009} (paper II) the underlying
theory is presented.

This paper is organized as follows. We start in Section II with a
quantum- chemical characterization of the molecule under
investigation. The experimental details are provided in Section III.
In Section IV two-dimensional electronic spectra are presented and
connections to the frequency-frequency correlation function are
established. We close in Section V with some concluding remarks. The
appendix contains a comparison to related one-dimensional four-wave
mixing signals.

\section{Quantum-chemical calculations}

This section characterizes PERY theoretically and spectroscopically with respect to its
two-dimensional $\pi$-conjugation by means of quantum-chemical
calculations. The electronic ground state geometry of PERY has been
optimized using Density Functional Theory (DFT) \cite{ParrYang1991}
based on the Becke's three parameter hybrid functional using the
Lee, Yang and Parr correlation functional (B3LYP) \cite{Becke1996}.
The polarized split-valence SV(P) basis set \cite{DunningHay1977}
has been used. The obtained optimal structures were checked by
normal mode analysis (no imaginary frequencies for all optimal
geometries). Based on the optimized geometry, the vertical
transition energies and oscillator strengths between the
initial and final states have been calculated using the
Configuration Interaction Singles (CIS) method \cite{ForesmanFrisch1992}
based on the semiempirical ZINDO/S (Zerner's Intermediate
Neglect of Differential Overlap) Hamiltonian \cite{ZernerWesterhoff1980}.
For the ZINDO/S calculations, the single excitations from the
10 highest occupied to the 10 lowest unoccupied molecular
orbitals were considered. The DFT calculations were done using the
Turbomole 5.7 package \cite{AhlrichsKoelmel1989} and the optical
transitions were computed using the Argus Lab software \cite{Thompson2004}.  The
evaluated theoretical values are used for the interpretation of
experimental spectroscopic measurements.

\begin{figure}
\begin{center}
\includegraphics{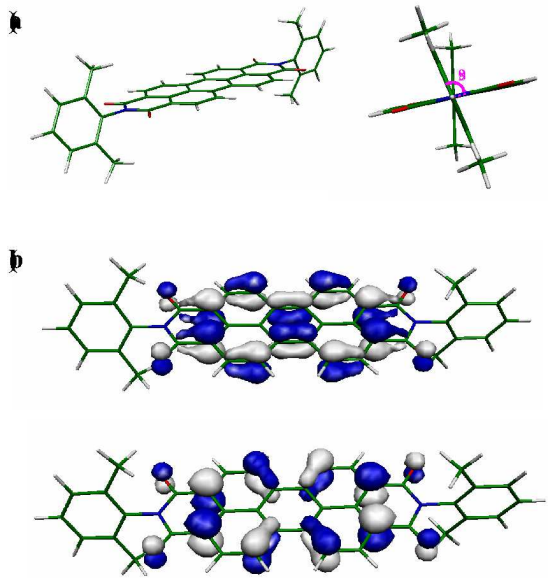}
\caption{(a) Side and front views on the B3LYP/SV(P) optimal structure
of PERY in the electronic ground state. (b)
Visualization of the HOMO (bottom) and LUMO (top) obtained
from ZINDO/S//B3LYP/SV(P) calculations.}\label{Fig_Qchem}
\end{center}
\end{figure}

The optimal geometry of D$_2$ symmetry exhibits a planar central
part (Fig.~\ref{Fig_Qchem}a). The orientation of the lateral
aromatic rings and the core skeleton is nearly perpendicular
(dihedral angles of 81.2$^{\circ}$) causing a negligible
$\pi$-conjugation between these molecular fragments. The lateral
rings are sterically hindered due to the presence of methyl groups in
ortho-position. The opposite mutual arrangement of the neighboring
lateral rings is not parallel as can be perceived from the right
part of Fig.~\ref{Fig_Qchem}a.

The photophysical properties of PERY are primarily determined by the
planar geometry of the central part and the related size of the
$\pi$-conjugation. The experimental absorption spectrum of PERY
in toluene (Fig.~\ref{Fig_LA}a) can
be characterized by an intense absorption band with four peaks which
are detected in the spectral range of 18000 to 24000~cm$^{-1}$. The equidistant energy difference
(1400~cm$^{-1}$) between the four peaks indicates a harmonic character of
the vibrational progression. The calculated vertical ZINDO/S
excitation energies and the oscillator strengths are in good
agreement with the experimental absorption spectrum. The ZINDO/S
method was parametrized for spectroscopic properties of the
$\pi$-conjugated aromatic systems and gives transition energies of
18725~cm$^{-1}$ with an oscillator strength f$=1.7814$, 28631~cm$^{-1}$
(f$=0.0001$), 31929~cm$^{-1}$ (f$=0.0002$), and 34536~cm$^{-1}$
(f$=0.8228$). The lowest energy transition is depicted as red
line in Fig.~\ref{Fig_LA}a. The
absence of other intensive electronic transitions in the region between
18000 and 24000~cm$^{-1}$ indicates the vibronic origin of the
progression observed in the experimental absorption spectrum.
Therefore, the first and most intensive peak in the experimental
record can be attributed to the $0\rightarrow 0$ vibronic transition. In
this context it is useful to examine the ZINDO/S HOMO and LUMO
of the molecule. The HOMO-LUMO excitation
plays a dominant role in this energetically lowest optical
transition according to the frontier molecular orbital theory, with its
contribution being more than 30\%. These selected
ZINDO/S orbitals are visualized in Fig.~\ref{Fig_Qchem}b,
demonstrating that both orbitals are delocalized over the center of
the core part only. The lobes of the LUMO are oriented more or less
perpendicular to the lobes of the HOMO. The shape
of the LUMO compared to those of the HOMO indicates
that the optical excitation is spread from the central atoms of the
core towards the nitrogen atoms.

In order to determine the origin of the vibronic motion reflected in
the linear and nonlinear spectroscopic experiments, vibrational line
spectra have been calculated. The complete set of computed
vibrational frequencies and intensities together with a
visualisation of the most dominant modes and Huang-Rhys factors are
collected in the Supporting Information. For determining the
Huang-Rhys factors semiemprical AM1 calculations were performed
using the Mopac program package 2002 \cite{Mopac2002}. The
calculated vibrational spectrum possesses 210 fundamentals between
13 and 3243~cm$^{-1}$. With the limited bandwidth of our excitation
pulses we can only cover modes up to frequencies of approximately
900~cm$^{-1}$. From Fig.~\ref{Fig_LA}b, in which modes with a
Huang-Rhys factor larger than 0.005 are plotted, it becomes obvious
that only a few modes contribute significantly to the observed
spectroscopic response.

\section{Experimental}

PERY was purchased from Sigma-Aldrich and used without further
purification. Solutions with a concentration of $3\cdot 10^{-4}$~M
were prepared by dissolving PERY in toluene (spectrophotometric
grade, Merck), sonication for $\approx$10~minutes, and filtration to
remove undissolved residues.

Pulse generation is accomplished by a regenerative titanium-sapphire
amplifier system (RegA 9050, Coherent Inc.) operating at a repetition
rate of 200~kHz. Conversion into
the visible spectral region is attained with a noncollinear optical
parametric amplifier (NOPA) \cite{PielEichberger2006}. For our present study the central
excitation energy was set to 18800~cm$^{-1}$ (FWHM 920~cm$^{-1}$)
and the pulses were attenuated to 10~nJ per pulse at the
sample position. Second-order dispersion is eliminated with a
sequence of fused-silica prisms, while chirped mirrors compensate
for third-order dispersion, yielding a pulse duration of 16~fs.
The pulse duration is determined with an
improved version of spectral phase interferometry for direct
electric field reconstruction (SPIDER), zero additional phase (ZAP)
SPIDER \cite{BaumRiedle2005}, and cross-checked with SHG-FROG
\cite{TrebinoKane1997}.

\begin{figure}
\begin{center}
\includegraphics{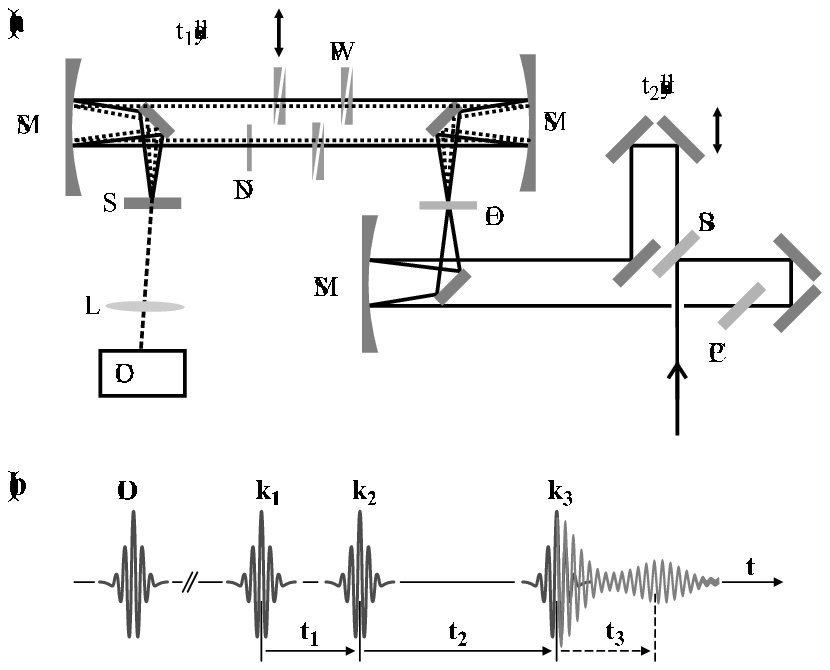}
\caption{(a) Schematic setup for recording 2D electronic spectra. BS
$\ldots$ beamsplitter, CP $\ldots$ compensation plate, SM $\ldots$
spherical mirror, DOE $\ldots$ diffractive optical element, WP $\ldots$ wedge pair, ND
$\ldots$ neutral density filter, S $\ldots$ sample, L $\ldots$ lens, CCD $\ldots$
charge coupled device camera. (b) Designation of pulses and time delays in four wave
mixing experiments.}\label{Fig_Setup}
\end{center}
\end{figure}

Our approach for recording 2D electronic spectra is based on a diffractive-optics
based setup following the experimental configuration reported in
Ref. \cite{CowanMiller2004, BrixnerFleming2005, MoranScherer2006}
(Fig.~\ref{Fig_Setup}a). A 50/50 beam splitter (BS) splits the NOPA output
into two beams of equal intensity, one of which can be delayed with
respect to the other by a computer controlled delay stage to
introduce delay $t_2$ (cf. Fig.~\ref{Fig_Setup}b for a definition
of time delays). A diffractive optical element (DOE), optimized for diffraction into $\pm$first
order, is used to split the two beams focused onto it into four beams
\cite{CowanMiller2004}. From this point on, all four beams are
reflected from common mirrors to maintain the phase correlations
\cite{CowanMiller2004, BrixnerFlemingI2004}. A spherical mirror (SM)
after the transmission grating parallelizes the four beams which are
arranged to form the four corners of a square. Pulse $\mathbf{k_2}$ passes
through a pair of glass wedges (WP) that are oriented in an anti-parallel
fashion \cite{BrixnerFlemingI2004}. By moving one of the wedges with
a computer controlled delay stage, the amount of glass in the beam
path and therefore the delay of the pulse is changed without
introducing lateral beam shifts. This configuration allows for
precise delay movements (5.3~as) without influencing the phase
stability of the phase locked pulse pair and is used to introduce delay $t_1$.
Pulses $\mathbf{k_1}$ and $\mathbf{k_3}$ pass
through equivalent but static wedge pairs to balance the dispersion. The LO is attenuated by a neutral density
filter (ND) by four to five orders of magnitude and reaches the sample
approximately 550~fs before the other pulses. All four beams are
focused by a spherical mirror onto the sample to generate the nonlinear signal. In order to
avoid unwanted signals from a cuvette window and to prevent
precipitation of the sample on the cell windows, a wire-guided
gravity driven jet \cite{TauberBradforth2003, LaimgruberGilch2006}
is implemented, yielding a film of approximately 200~$\mu$m
thickness. Due to the square geometry
of the excitation beams and the LO, the signal propagation direction
is coincident with the propagation direction of the LO allowing for
heterodyne detection of the signal field with the LO. Data acquisition is accomplished by means of a
thermoelectrical cooled charge-coupled-device (CCD) spectrometer. A
2400~lines/mm grating in combination with a 1024~pixel array allows
for a spectral resolution of 3.3~cm$^{-1}$ in $\omega_3$.

For recording 2D electronic spectra, $t_2$ is fixed at a certain
value and $t_1$ is scanned over $\pm$100~fs with a step size of
0.65~fs to fulfill the Nyquist sampling criterion. The electric field
of the signal as a function of frequency ($\omega_3$) is
reconstructed by Fourier transform of each interferogram (for every
$t_1$-step), filtering and back Fourier transform
\cite{LepetitJoffreI1996}. Fourier transform along $t_1$ yields the
second frequency dimension $\omega_1$. The frequency resolution in
$\omega_1$ is solely determined by the maximal temporal separation
between pulses $\mathbf{k_1}$ and $\mathbf{k_2}$ and amounts to 26~cm$^{-1}$ in our
experiments. In a final step the absolute phase of the 2D spectrum has to be determined.
Since the PP signal does not depend
on constant phase shifts of either the pump or the probe beam, it
can be used for this purpose \cite{Jonas2003}. To this end the projection slice theorem is
applied, that states that the projection of the real part of the 2D
spectrum onto $\omega_3$ equals the spectrally resolved (SR) PP
signal. By multiplying this projection by a factor of e$^{-i\phi}$
and adjusting $\phi$ so that the projection matches the SRPP signal,
the absolute phase of the 2D spectrum is recovered.

\section{Results and Discussion}

\subsection{Two-dimensional electronic spectra}

With second-order nonlinear signals being symmetry-forbidden in isotropic
media such as bulk liquids, the third-order nonlinearity is the lowest
nonlinear order at which system-bath interactions and solute dynamics in liquids can
be probed. The third-order nonlinear polarization $P^{(3)}(t_1,t_2,t_3)$
is thereby induced by three interactions with pulsed electromagnetic fields
(with wavevectors $\mathbf{k_1}$, $\mathbf{k_2}$, $\mathbf{k_3}$ and temporal
separation $t_1$ and $t_2$)
and it radiates the signal in the phase-matched directions $\mathbf{\pm k_1\pm k_2\pm k_3}$.
Depending on the scanning procedure, the detection direction (phase
matching), and the detection scheme (time vs. frequency resolved;
homodyne vs. heterodyne detection), different techniques can be distinguished. In this subsection we focus
on two-dimensional electronic spectroscopy, related one-dimensional four-wave
mixing techniques are discussed in the Appendix.

\begin{figure*}
\begin{center}
\includegraphics[scale=0.9]{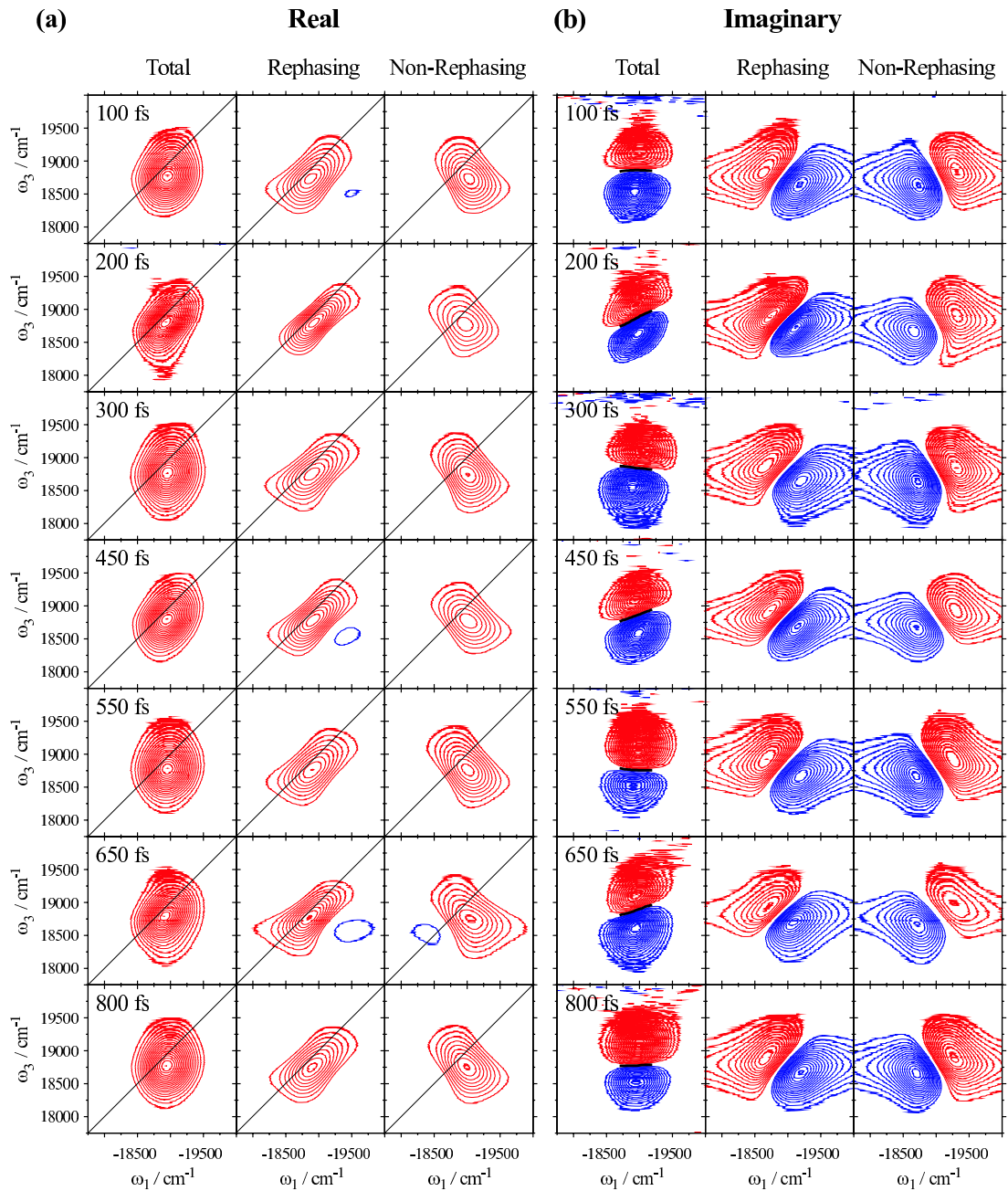}
\caption{(a) Real and (b) imaginary part of the 2D spectra of PERY
in toluene for $t_2$-delays of 100, 200, 300, 450, 550,
650, and 800~fs. The first, second, and third column of each panel
display the corresponding total signal, the rephasing, and the
non-rephasing part, respectively. All total spectra are normalized
to their absolute maximum value, whereas the rephasing and
non-rephasing parts are plotted with their respective contribution
to the total signal. Contour lines are drawn at 5$\%$ intervals
starting at $\pm 10\%$. Red lines indicate positive signals, blue
lines negative ones. The solid line in the real part indicates the
diagonal, whereas the black lines in the imaginary part are drawn at
the zero crossings between the positive and negative
features.}\label{Fig_2D_Re_Im}
\end{center}
\end{figure*}

\begin{figure*}
\begin{center}
\includegraphics[scale=0.9]{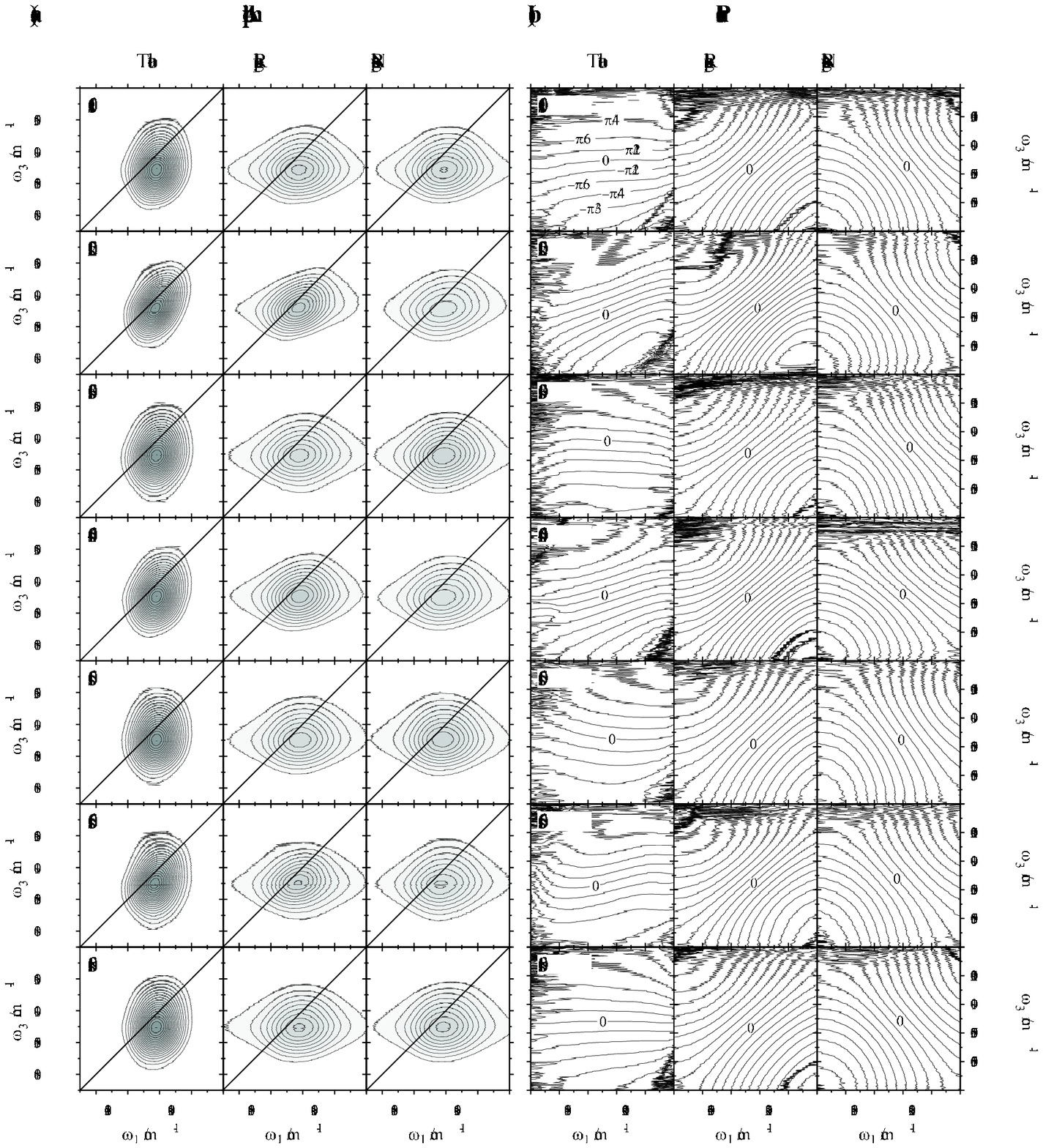}
\caption{(a) Amplitude and (b) phase of the 2D spectra of PERY in
toluene recorded at $t_2$-delays of 100, 200, 300, 450,
550, 650, and 800~fs. The first, second, and third column of each
panel display the corresponding total signal, the rephasing and the
non-rephasing part, respectively. The amplitude part total spectra
are normalized to their absolute maximum value, whereas the
rephasing and non-rephasing parts are plotted with their respective
contribution to the total signal. Contour lines are drawn at 5$\%$
intervals starting at $\pm 10\%$ in the amplitude part spectra and
at $\pm\frac{\pi}{12}$ intervals in the phase spectra. The
zero-phase line is indicated in each phase spectrum. Note the
different spectral range in the representation of the phase
spectra.}\label{Fig_2D_Am_Phase}
\end{center}
\end{figure*}

In contrast to one-dimensional FWM techniques, where time- or
frequency-integrating detection schemes project the full information
content onto a single axis (i.e. intensity vs. time or frequency),
2D spectra provide a more complete picture by extending the
dimensionality of representation. Selective excitation,
competing with high temporal resolution in one-dimensional
experiments, is not an issue in 2D spectroscopy owing to the
mathematical properties of Fourier transform, i.e. the frequency
resolution being determined by the scanned time range and vice
versa. 2D electronic spectra are recorded by scanning $t_1$ over a
defined interval for a fixed value of $t_2$ and recording the signal
in the phase matched directions $\mathbf{\pm k_1\mp k_2+k_3}$ in a
heterodyne detection scheme.
Heterodyne detection serves not only to
boost weak signals, but also to extract the amplitude and phase of
the signal field by Fourier transform algorithms
\cite{HyblJonas1998, LepetitJoffreI1996}.
Fourier transform with respect to $t_1$ and $t_3$ results in a
complex signal of the form

$$
S_{2D}(\omega_1, t_2, \omega_3) =
\int_0^{\infty}\int_0^{\infty}iP^{(3)}(t_1,t_2,t_3)
$$
\begin{equation}
\times\textrm{e}^{+i\omega_3t_3}\textrm{e}^{\pm
i\omega_1t_1}\textrm{d}t_1\textrm{d}t_3.
\end{equation}

\noindent where the negative sign in the exponent accounts for the rephasing
signal part (i.e. $t_1\geq 0$, $\mathbf{k_s=-k_1+k_2+k_3}$) and the positive sign
for the non-rephasing signal part (i.e. $t_1\leq 0$, $\mathbf{k_s=+k_1-k_2+k_3}$).
Complex signals can be
represented either by their real and imaginary parts or by their
amplitude and phase. These two representation schemes are displayed
for PERY in Figs.~\ref{Fig_2D_Re_Im} and
\ref{Fig_2D_Am_Phase} for $t_2$-times of 100, 200,
300, 450, 550, 650, and 800~fs.

The real part of a 2D spectrum (first column in
Fig.~\ref{Fig_2D_Re_Im}a) is the product of absorptive line-shapes
in $\omega_1$ and in $\omega_3$. In agreement with approximating PERY as a two-level electronic system,
where ground state bleaching and stimulated emission are the only
two contributions, it shows a single positive feature. The maximum
of the peak is centered slightly below the diagonal
($|\omega_1|>\omega_3$), indicative of a small Stokes shift. The
peak experiences periodic modulations in its shape with a period
of 240~fs (a detailed analysis is presented in Fig.~\ref{Fig_Oszillation}). For
$t_2$-times of 200, 450, and 650~fs, the shape of the absorptive
peak is strongly elliptical with the major axis of the ellipse
oriented along the diagonal. Contrarily, for $t_2=100, 300, 550$, and 800~fs the
absorptive peak adopts a more circular shape and the major axis is
oriented nearly parallel to the $\omega_3$-axis.

The imaginary part of a 2D spectrum is the product of an absorptive
line-shape in $\omega_1$ and a dispersive line-shape in $\omega_3$.
In case of a two-level system the imaginary part features two
contributions of opposite sign (first column in
Fig.~\ref{Fig_2D_Re_Im}b). Similar to the real part, the imaginary
part too exhibits oscillations with a period of 240~fs. The most obvious oscillation
is seen in the nodal line separating the positive and negative
feature. The angle of this nodal line with respect to the
$\omega_1$-axis changes from being positive for $t_2=200$, 450, and
650~fs, to zero or even negative values for $t_2=100$, 300, 550, and
800~fs (cf. Fig.~\ref{Fig_Oszillation} for a detailed evaluation).
In addition to this oscillation of the nodal line, the width of both
contributions periodically changes from being narrow for $t_2=200$,
450, and 650~fs, to being broad for $t_2=100$, 300, 550, and 800~fs.

In Fig.~\ref{Fig_2D_Am_Phase} we show amplitude and phase 2D spectra
of PERY for $t_2$-delays of 100, 200, 300, 450,
550, 650, and 800~fs. The amplitude spectra (first column in
Fig.~\ref{Fig_2D_Am_Phase}a) resemble the real part spectra shown in
Fig.~\ref{Fig_2D_Re_Im}a, with a single feature centered
slightly below the diagonal. Similar to the real part spectra this
positive peak features oscillations in its orientation and shape.
For $t_2=$ 200, 450, and
650~fs the peak shape is highly elliptical and its major axis is
oriented along the diagonal. For $t_2=$ 100, 300, 550, and 800~fs the peak acquires a more
circular shape and is oriented nearly parallel to the
$\omega_3$-axis.

In the phase spectra (first column in Fig.~\ref{Fig_2D_Am_Phase}b)
the lines of constant phase exhibit oscillations in their
orientation. For $t_2=$ 200, 450, and 650~fs the lines of constant
phase are oriented along the diagonal, whereas they turn to become
horizontal at $t_2=$ 100, 300, 550, and 800~fs. Note that the phase
spectra are shown for a narrower range than the real, imaginary, and
amplitude part spectra. All the observed line-shape modulations will
be assigned below to periodic modulations in the strength of the
rephasing and non-rephasing contributions to the total signal.

\subsection{Rephasing and non-rephasing signal parts}

\begin{figure}
\begin{center}
\includegraphics{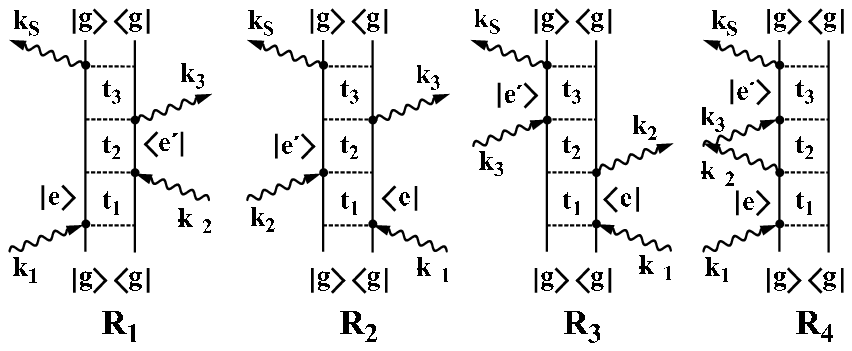}
\caption{The four Feynman diagrams graphically
illustrating the Liouville space pathways for a two-level system
within the rotating wave approximation. \emph{g} and \emph{e} denote ground and
excited state, respectively.}\label{Fig_Feynman}
\end{center}
\end{figure}

The third-order nonlinear polarization $P^{(3)}(t_1,t_2,t_3)$, which
underlies all FWM signals, can be
expressed as a convolution of the appropriate response functions
$R_i(t_1, t_2, t_3)$ with the electric fields of the excitation
pulses \cite{Mukamel1995}. The four Feynman
diagrams graphically illustrating the four response functions for a
two-level system within the rotating wave approximation are depicted
in Fig.~\ref{Fig_Feynman}. In these diagrams the
left and right vertical lines indicate the evolution of the
$|ket\rangle$ and $\langle bra|$ of the density matrix,
respectively. Wavy arrows denote interactions with the
electromagnetic field, and time evolves from bottom to top. In all
four diagrams, the first interaction of the laser field with the
molecular system induces a coherence between the ground state ($g$)
and the first excited state ($e$) in which the system evolves for a time $t_1$.
Interaction with the
second laser pulse creates a population or vibrational coherence in
either the ground (diagrams $R_3$ and $R_4$) or the excited
(diagrams $R_1$ and $R_2$) state. After a time $t_2$ the third pulse again induces a
coherence between the ground and excited state that radiates the
signal field in the phase-matched direction $\mathbf{k_s}$.

The four Feynman-diagrams shown in Fig.~\ref{Fig_Feynman} can be classified
as ground state bleaching (GSB) and stimulated emission (SE), depending on
whether the system evolves in the ground state ($R_3$, $R_4$)
or in the excited state ($R_1$, $R_2$) during $t_2$, respectively.
A different classification of these four diagrams arises from the
ordering of the first two interactions. In diagrams $R_2$ and
$R_3$ the first interaction takes place with pulse
$\mathbf{-k_1}$, whereas in diagrams $R_1$ and $R_4$ the first
interaction takes place with positive sign ($\mathbf{+k_1}$). As a
consequence the system evolves in conjugate frequencies during
$t_1$ ($\omega_{ge}$) and $t_3$ ($\omega_{eg}$) in diagrams $R_2$
and $R_3$. This allows for rephasing in an inhomogeneously
broadened ensemble, resulting in the formation of an "echo" in
case the system has kept some memory of its transition frequency
over $t_2$. These diagrams therefore contribute to the rephasing
signal part. In contrast, the system evolves with the same
frequency during $t_1$ and $t_3$ ($\omega_{eg}$) in diagrams $R_1$
and $R_4$. These two diagrams are denoted non-rephasing and the
systems's response is close to a free induction decay. Rephasing
diagrams contribute to the signal for positive delays $t_1$ (i.e.
pulse 1 preceding pulse 2), whereas non-rephasing diagrams
contribute for negative delays of $t_1$ (i.e. pulse 2 preceding
pulse 1). In addition to these four diagrams there are diagrams
that contribute to the signal only if the pulse separation is
shorter than the pulse duration, i.e. when the excitation pulses
overlap. These diagrams will not be taken into account in our
discussion.

Scanning $t_1$ over positive and negative values equally weights
rephasing and non-rephasing pathways. This scanning procedure has
been shown to produce purely absorptive real parts and purely
dispersive imaginary parts, that are linked via the Kramers-Kronig
relation \cite{HyblJonasI2001, KhalilTokmakoffI2003}. The
rephasing and non-rephasing spectra on their own exhibit
phase-twisted line-shapes resulting from mixing of dispersive and
absorptive features. The dissection of the total 2D spectrum into
its corresponding rephasing and non-rephasing parts is shown in
the second and third row of Figs.~\ref{Fig_2D_Re_Im} and
\ref{Fig_2D_Am_Phase} for the real part, imaginary part,
amplitude, and phase, respectively. The decomposition was
performed by recording the full scan and evaluating only positive
$t_1$-values for the rephasing part and only negative $t_1$-values
for the non-rephasing part \cite{Remark1}. Strictly speaking,
rephasing signals show up at \{$-\omega_1, +\omega_3$\} according
to the conjugate frequency evolution during $t_1$ and $t_3$.
Non-rephasing signals on the other hand appear at \{$+\omega_1,
+\omega_3$\}, since the system evolves with the same frequency
during $t_1$ and $t_3$. However, to facilitate comparison of the
rephasing, non-rephasing, and total spectra, we plot all signals
along \{$-\omega_1, +\omega_3$\} \cite{GeHochstrasser2002,
KhalilTokmakoffII2003}.

From the inspection of Fig.~\ref{Fig_2D_Re_Im}a, one can immediately
see that both rephasing and non-rephasing parts of the real part 2D
spectra are of a pronounced elliptical shape. However, the major
axis of the ellipse in case of the rephasing part is oriented along
the diagonal, whereas it is oriented along the anti-diagonal in case
of the non-rephasing part. Each part on its own exhibits
oscillations of its amplitude without significant line-shape
modulations. The amplitude oscillations, however, are out of phase
by $\pi$, i.e. the rephasing part has its maxima at $t_2$-times of
200, 450 and 650~fs, whereas the non-rephasing part has its maximal
amplitude at $t_2=100$, 300, 550, and 800~fs. Therefore, for
$t_2$-times of 200, 450, and 650~fs the major contribution to
the total signal stems from the rephasing part, whereas for $t_2$-times
of 100, 300, 550, and 800~fs the non-rephasing part
contributes more strongly to the total spectrum. In
Fig.~\ref{Fig_Oszillation} we present a quantitative
evaluation of the respective contributions.

An analogous situation is encountered in the imaginary part as shown
in Fig.~\ref{Fig_2D_Re_Im}b. The nodal line of the rephasing part is
oriented along the diagonal, whereas the one of the non-rephasing
part is oriented along the anti-diagonal. As a consequence of the
out-of-phase oscillations of the amplitudes of the two
contributions, the nodal line of the total signal experiences a
periodic modulation of its orientation. Similar to the real part,
the rephasing and non-rephasing imaginary peak shapes do not change
significantly with increasing $t_2$-time, it is only their relative
contribution to the total signal that varies with $t_2$. From
inspection of the dispersive part it becomes most obvious how the
phase-twisted line-shapes in the rephasing and non-rephasing signal
parts cancel upon summation to yield a purely dispersive imaginary
part. In the accompanying paper II we explain the oscillations of the
rephasing, nonrephasing, and sum spectra based on an expansion of the
frequency-frequency correlation function in terms of Huang-Rhys factors.

The dissection into rephasing and non-rephasing parts was also
performed for the amplitude and phase spectra
(Fig.~\ref{Fig_2D_Am_Phase}). For these representations, however, simple
addition of the two contributions to yield the total spectrum is
not possible. Nevertheless, the oscillations in the total spectra
can be assigned to oscillating features in the rephasing and
non-rephasing signal parts. In the amplitude part 2D spectra
(Fig.~\ref{Fig_2D_Am_Phase}a), we observe a stronger contribution
of the rephasing part for $t_2$-delays of 200, 450, and 650~fs,
and a reverse situation (i.e. an intensity gain of the
non-rephasing contribution) for $t_2$-delays of 100, 300, 550, and
800~fs. Therefore, the total signal is elliptical and
line-narrowed along the anti-diagonal in the former case, and more
circular in the latter. In the phase spectra
(Fig.~\ref{Fig_2D_Am_Phase}b), the lines of constant phase are
oriented along the diagonal in the rephasing part and along the
anti-diagonal in the non-rephasing part. This behavior reflects
the different sign of the phase accumulated during the two
time-intervals $t_1$ and $t_3$ according to the different
coherence evolution in the rephasing and non-rephasing Feynman
diagrams \cite{LoparoTokmakoff2006}. Depending on the relative
strength of the rephasing and non-rephasing contributions, lines
of constant phase in the total phase spectrum are either oriented
along the diagonal ($t_2$=200, 450, and 650~fs), or they rotate
down to become approximately parallel to the $\omega_1$-axis
($t_2$=100, 300, 550, and 800~fs). In the next section we discuss
relations of the line-shape evolutions in 2D electronic spectra to
the frequency-frequency correlation function.

\subsection{Frequency-frequency correlation function}

For electronic spectra in the condensed phase, where transition
energy fluctuations are non-Markovian in nature, homogeneous and
inhomogenous timescales are not readily separable. Therefore, such
systems are usually described by a frequency-frequency
correlation function (FFCF) $M(t)$, defined as

\begin{equation}
M(t)=\frac{\langle\delta\omega_{eg}(t)\delta\omega_{eg}(0)\rangle}{\langle\delta\omega_{eg}^2\rangle}.
\end{equation}

$M(t)$ is a normalized ensemble averaged product of transition
frequency fluctuations ($\delta\omega_{eg}$) separated by time
$t$. Since the correlation function in principle carries all
relevant information on intramolecular and system-bath dynamics,
considerable attention has been drawn over the last decade on how
to sample the FFCF by experimentally feasible methods
\cite{deBoeijWiersmaII1996, JooFleming1996, deBoeijWiersmaI1998}.
Once $M(t)$ is known, all linear and non-linear spectroscopic
signals can be calculated via the line-shape function $g(t)$
with knowledge of the reorganization energy $\lambda$ and the
temperature dependent coupling parameter, which reduces to
$\frac{2\lambda k_B T}{\hbar}$ at high temperatures. We use
such complete information about the transition frequency fluctuations
due to intramolecular and system bath coupling to derive a detailed line
shape theory in Paper II \cite{MancalSperling2009}. Here, we concentrate
on the intramolecular part of the $M(t)$-function which is responsible
for the oscillatory dynamics of the line shapes.

Although signal detection of one-dimensional FWM signals
is relatively straightforward, these methods face an inherent
trade-off between time- and frequency-resolution. This obstacle is
evaded in 2D-ES, where model calculations and experiments in the
IR and VIS spectral region revealed correlations between
line-shape evolutions and the frequency-frequency correlation
function \cite{HyblJonasII2001, WoutersenStock2002,
KhalilTokmakoffII2003, KwacCho2003, RobertsTokmakoff2006,
LazonderWiersma2006, LoparoTokmakoff2006}.

\begin{figure}
\begin{center}
\includegraphics{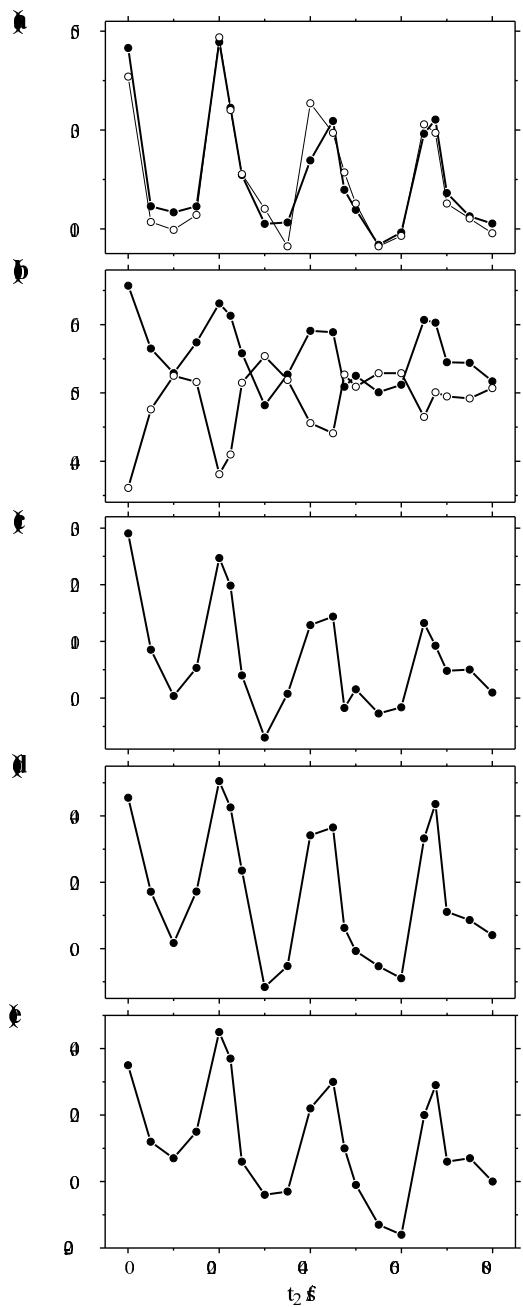}
\caption{Oscillating features extracted from the 2D spectra of PERY
in toluene. (a) Ellipticity (filled circles) and center line
slope (open circles) of the peak in the real 2D
spectra. (b) Relative amplitudes of
the rephasing (filled circles) and the non-rephasing (open circles) part of the
real 2D spectra. (c) Inhomogeneity
index extracted from the real part 2D spectra as defined in
Eq.~\ref{Eq_inhomogeneityindex}. (d) Slope of the nodal line
separating the positive and negative feature in the imaginary 2D
spectra. (e) Slope of the $\phi=0$
phase line in the phase 2D spectra.}\label{Fig_Oszillation}
\end{center}
\end{figure}

Two quantities extracted from the real part of 2D electronic
spectra were shown to be directly proportional to the FFCF: the
ellipticity of the peak shape \cite{LazonderWiersma2006} and the
center line slope \cite{KwakFayer2007, ParkFayer2007}. The
ellipticity of the peak as a function of $t_2$ is defined as
\cite{RobertsTokmakoff2006, LazonderWiersma2006}

\begin{equation}\label{Eq_real}
E(t_2)=\frac{D^2-A^2}{D^2+A^2},
\end{equation}

\noindent where D and A denote the diagonal and anti-diagonal peak
widths, respectively. In Fig.~\ref{Fig_Oszillation}a we plot this
ratio for $0<t_2<800$~fs, with the diagonal and anti-diagonal peak
widths evaluated at $\frac{1}{e}$-height. Oscillations with a period
of 240~fs are clearly recovered. In our previous
work we have shown that these oscillations arise mainly from
modulations of the anti-diagonal width, whereas the diagonal width
stays nearly constant over the time-range explored in our
experiments \cite{NemethSperlingI2008}. The center line slope
(CLS) is extracted by taking cuts parallel to the $\omega_3$-axis
and finding the maximum of the peak. The resulting data points as
a function of $\omega_1$ are fit to a first order polynomial. The
slope of this line is the center line slope
(Fig.~\ref{Fig_Oszillation}a). Note that in the work by Kwak
\emph{et al.} \cite{KwakFayer2007} cuts are taken parallel to the $\omega_1$-axis and
the CLS is defined as the inverse of the slope of the line
connecting the maxima of the slices. Both the ellipticity and the
CLS can vary between 1 and 0 and directly reflect the
$t_2$-dependent part of the FFCF \cite{LazonderWiersma2006,
KwakFayer2007} . The limiting value of 1 can be reached for
$t_2=0$ and in the absence of a homogeneous component. The
deviation from the initial value of 1 can thereby be related to
the magnitude of the homogeneous component, as was shown in Ref.~\cite{KwakFayer2007}. Relaxation
and spectral diffusion processes during $t_2$ lead to a decrease
of the ellipticity and the CLS, until in the long time limit both
quantities approach the value of zero, indicative of equal widths
along the diagonal and anti-diagonal directions and a slope of the
center line that is parallel to the $\omega_1$-axis. In a previous
work it was argued that the CLS has the advantage over the
ellipticity of being independent of factors influencing the
line-shape (e.g. finite pulse duration, apodization, etc.) \cite{KwakFayer2007}. As can
be seen in Fig.~\ref{Fig_Oszillation}a we do not observe
significant differences between these two quantities. Closely
related to the ellipticity of the peak shape is the eccentricity,
which is defined as $Ec=\sqrt{1-\frac{A^2}{D^2}}$ and which has
been proposed in earlier works as a measure for the FFCF
\cite{FinkelsteinFayer2007}. Even though the eccentricity takes
the same limiting values as the ellipticity, i.e. 1 if D$>>$A and
0 if D$=$A (which is the case for a circle), the two quantities
differ for intermediate cases and therefore cannot be compared
directly.

\begin{figure*}[ht]
\begin{center}
\includegraphics[scale=0.9]{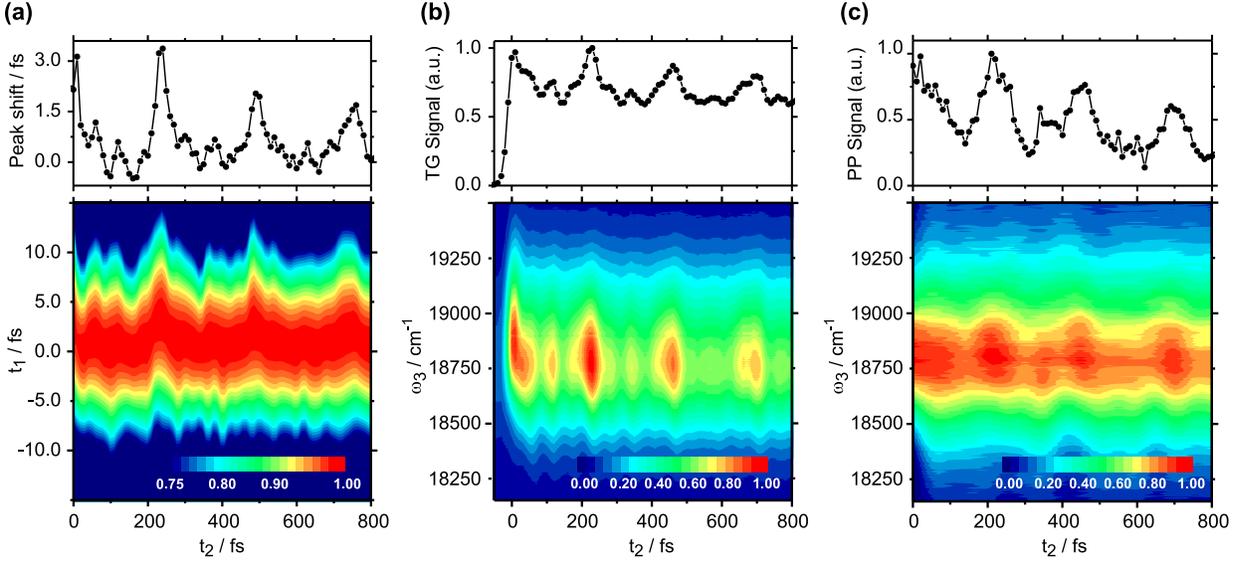}
\caption{One-dimensional four-wave mixing signals of PERY in
toluene. Top and bottom panels share the same abscissa, color
scales are shown as insets in the bottom panels. (a) Three-pulse
photon echo peak-shift trace (top) and the corresponding
frequency integrated three-pulse photon echo signal as a
function of delay $t_1$ and $t_2$. (b) Frequency integrated
(top) and frequency resolved (bottom) transient grating signal.
(c) Frequency integrated (top) and frequency resolved (bottom)
pump-probe signal. }\label{Fig_3PEPS_TG_PP}
\end{center}
\end{figure*}

As discussed above, a periodic modulation of the rephasing and
non-rephasing contribution to the total 2D spectrum induces the
observed changes in the real and imaginary parts. In
Fig.~\ref{Fig_Oszillation}b we plot the relative amplitudes of the
rephasing and non-rephasing part of the real 2D spectra. We
clearly observe out-of-phase oscillations of these two
contributions, with the rephasing part exhibiting its
maxima at $t_2$-times of 200, 450, and 650~fs and its minima
at $t_2=$ 100, 300, 550, and 800~fs, and the non-rephasing part showing the opposite behavior,
i.e. having its maxima at $t_2$-times corresponding to minima in
the rephasing part and vice versa. Tokmakoff \emph{et al.}
\cite{LoparoTokmakoff2006, RobertsTokmakoff2006} defined an
inhomogeneity index $I(t_2)$ as

\begin{equation}\label{Eq_inhomogeneityindex}
I(t_2)=\frac{A_R-A_{NR}}{A_R+A_{NR}},
\end{equation}

\noindent with $A_R$ being the relative amplitude of the rephasing,
and $A_{NR}$ the relative amplitude of the non-rephasing part. This
quantity was shown to directly relate to $M(t)$ via
$M(t_2)=\sin{(\frac{\pi I}{2})}$. From the inspection of the
inhomogeneity index (Fig.~\ref{Fig_Oszillation}b) one can notice,
that the non-rephasing part exceeds the rephasing part in signal
strength around $t_2$-delays of 300 and 550~fs. For these delays the
inhomogeneity index adopts negative values, which is consistent with
negative peak shift values observed in Fig.~\ref{Fig_3PEPS_TG_PP}a
(cf. Appendix).

In the imaginary part of 2D spectra it is the slope of the nodal
line separating the positive and negative contribution that is
related to $M(t)$ \cite{KwacCho2003}. We extract this slope from
our 2D spectra by finding the zero crossing between the positive
and negative contribution in the range
$-18700$~cm$^{-1}<\omega_1<-19200$~cm$^{-1}$ and fitting of the
resulting data points with a first order polynomial. The slope of
the nodal line oscillates with a period of 240~fs (Fig.~\ref{Fig_Oszillation}d). Note that
also the slope takes negative values around $t_2=300$ and 550~fs.
This agrees well with the observation of negative values of the
inhomogeneity index (Fig.~\ref{Fig_Oszillation}c) and the peak
shift (Fig.~\ref{Fig_3PEPS_TG_PP}a). In case of a
three-level system it is the real part that features two peaks of
opposite sign, with the positive one stemming from ground state
bleaching and stimulated emission and the negative feature arising
from excited state absorption. In this case the slope of the nodal
line is extracted from the real part spectra
\cite{HyblJonasII2001, EavesGeissler2005, LoparoTokmakoff2006}.

Extracting the frequency-frequency correlation function from the
phase spectra (Fig.~\ref{Fig_2D_Am_Phase}b) is done by evaluating
the slope of lines of constant phase \cite{RobertsTokmakoff2006,
LoparoTokmakoff2006}. In Fig.~\ref{Fig_Oszillation}e we plot this
slope for $\phi=0$ evaluated in the frequency range
$-18500$cm$^{-1}<\omega_1<-19250$~cm$^{-1}$. Again, the
oscillating pattern exhibits a period of 240~fs with maxima at $t_2$-delays of
200, 450, and 650~fs and minima around $t_2=$100, 300, 550, and
800~fs. Around 300 and 550~fs the slope of lines of constant phase
adopts negative values indicative of a tilt towards the
anti-diagonal axis.

\section{Conclusion}

Intramolecular and system-bath (solvation) dynamics can be studied
either in time-domain by the frequency-frequency correlation
function $M(t)$, or, equivalently, in frequency-domain by the
spectral density $\rho(\omega)$. In this contribution we have
concentrated on time-domain measures, by presenting different
metrics that can be extracted from two-dimensional electronic
spectra and that follow the functional form of the
frequency-frequency correlation function. These metrics are the
ellipticity, the center line slope, and the inhomogeneity index
extracted from the real part spectra, the slope of the nodal line in
the imaginary part spectra, and the slope of lines of constant phase
in the phase spectra. Peak shift curves of PERY measured in solvents
of different polarity (nonpolar, polar, and weakly polar) featured
similar timescales \cite{LarsenFleming1999}, thus pointing to weak
solute-solvent interactions. The observed oscillations in the
frequency-frequency correlation function can thus be assumed to be
dominated by intramolecular dynamics. The observed beating pattern
stems from the fact that during $t_2$ the density matrix does not
only evolve in a population state (ground or excited state), but
vibrational coherences can be excited as well. These coherences
evolve with a frequency that is given by the difference in frequency
of the two states involved. In theoretical works on vibrational
effects in 2D-ES, the vibrational sub-levels could be resolved as a
consequence of the customized input parameters
\cite{GallagherJonas1999, EgorovaDomckeI2007}. Under experimentally
feasible conditions effects as finite pulse durations and finite
temperature erode the multipeak structure. In such a situation,
vibrational dynamics can not be followed by the evolution of
diagonal and cross peaks, but by the evolution of line-shapes and
signal intensities, as has been done in the present contribution. In
our previous work \cite{NemethSperlingI2008} we performed a fit to
the experimental two-dimensional spectra by including two overdamped
solvent modes and two underdamped intramolecular modes. In the
second paper of this series we investigate the influence of
vibrational coupling on 2D electronic spectra more thoroughly by
employing nonlinear response function theory and expanding the
frequency-frequency correlation function in terms of Huang-Rhys
factors. Furthermore, we extend the theory also to the case of high
frequency modes in which case individual diagonal and cross peaks
can be distinguished.

\vspace{1cm}
 {\bf Acknowledgements.} This work was supported by
the Austrian Science Foundation (FWF) within the projects No.
P18233 and F016/18 \textit{Advanced Light Sources} (ADLIS). A.N.
and J.S. thank the Austrian Academy of Sciences for partial
financial support by the Doctoral Scholarship Programs
(DOCfFORTE and DOC). T. M. acknowledges the kind support by the
Czech Science Foundation through grant GACR 202/07/P278 and by
the Ministry of Education, Youth, and Sports of the Czech
Republic through research plan MSM0021620835. J. H. gratefully
acknowledges support by the Lise Meitner project M1080-N16. The
quantum-chemical calculations were performed in part on the
Schr\"{o}dinger III cluster at the University of Vienna.

\vspace{1cm}
{\bf Supplemental Material.} The Cartesian
B3LYP/SV(P) coordinates, total electronic energy, harmonic
vibrational spectrum, Huang-Rhys factors, and a visualisation of
the most dominant vibrational modes of PERY can be found in the
supplemental material.

\appendix

\section{Comparison to other four-wave mixing signals}
Heterodyned two-dimensional electronic spectroscopy gathers the
maximum amount of information that can be inferred by any
third-order nonlinear technique. Nevertheless it is instructive to
cross-check our findings against results that are obtained with
alternative detection schemes, shown in Fig.~\ref{Fig_3PEPS_TG_PP}.
From the experimental point of view, these signals may be
differentiated by the required measuring (scanning) time and set-up
complexity, while, on the other hand, all of them are related to the
functional form of $M(t)$ \cite{deBoeijWiersmaI1996, JooFleming1996,
deBoeijWiersmaI1998}.

The three-pulse photon echo peak-shift (3PEPS) method is a variant
of homodyned photon echo spectroscopy, in which the time-integrated
nonlinear signal

\begin{equation}
S_{3PE}(t_1,
t_2)=\int_0^{\infty}|P^{(3)}(t_1,t_2,t_3)|^2\textrm{d}t_3
\end{equation}

\noindent is recorded in a wavevector architecture equivalent to 2D-ES. In
each step of a 3PEPS sequence, delay $t_1$ is scanned at a fixed
delay $t_2$, and the position of the (time- and
frequency-integrated) signal maximum evaluated along the $t_1$-axis.
The magnitude of this peak-shift (deviation from $t_1=0$) is
recorded for a range of $t_2$-delays, to give the peak-shift decay.
In essence, a finite photon echo peak-shift value indicates the
system's ability to rephase a macroscopic coherence (i.e. to cause
an echo) after having spent a certain time ($t_2$) in an
intermediate state. Fig.~\ref{Fig_3PEPS_TG_PP}a shows the photon
echo peak-shift decay for PERY in toluene, along with a
two-dimensional plot of $S_{3PE}(t_1, t_2)$. Both data
representations clearly reveal how the photon echo signal maximum
oscillates (along $t_1$) as a function of $t_2$, the oscillation
being observable for $t_2$-delays well beyond one picosecond. In the
context of our analysis of the 2D electronic spectra of PERY, it is
interesting to note that the peak-shift acquires negative values
around the minima of the peak-shift trace (located at $t_2$-delays
around 100, 350, and 600~fs). In other words, the integrated photon
echo (recorded as a function of $t_1$) may peak at negative
$t_1$-delays, if non-rephasing signal contributions dominate over
rephasing contributions to the total signal. We point out that this
finding is consistent with the negative value for the inhomogeneity
index and a negative slope of the nodal line in the dispersive 2D
signal part, observed at the corresponding $t_2$-delays.

For the sake of completeness, Fig.~\ref{Fig_3PEPS_TG_PP}b and c show
the transient grating (TG) and the pump-probe (PP) signals of PERY,
in both frequency integrated and frequency resolved representation.
The former method records $S_{3PE}$ as a function of $t_2$ for delay
$t_1$ set to zero (descriptively, the first two excitation pulses
form a spatial transmission grating from which the third pulse is
scattered off). Note that unlike 3PEPS, TG measurements are
sensitive to population decay during $t_2$. Time delays are scanned
in a similar way in PP measurements (i.e. scanning $t_2$ for
$t_1=0$), however, only two pulses are involved in the experiment,
the signal being recorded along the direction of the second pulse.
Since the PP signal is generated by two interactions with the first
pulse (pump), followed by one interaction with the second pulse
(probe), $t_1$ is intrinsically zero. Overall, also our TG and PP
data-sets do well align with previously published data
\cite{LarsenFleming1999, LarsenFleming2001} and our 2D electronic
spectra. In both types of signals, strong intensity oscillations
with a period of 240~fs (140~cm$^{-1}$) can be discerned. The slow
decay of the TG signal confirms the absence of any fast population
decay channels. In line with the discussion above, the PP signal is
dominated by (positive) stimulated emission and ground state
bleaching contributions, indicating excited state absorption effects
to be negligible on the picosecond timescale investigated here.

\end{document}